\RequirePackage{fix-cm}

\documentclass[smallextended]{svjour3}       
\smartqed  
\usepackage{graphicx}
\usepackage{subfigure}
\usepackage{textcomp}
\usepackage{gensymb}
\usepackage{amsmath}
\usepackage{enumitem,amssymb}
\usepackage[authoryear]{natbib}  
\usepackage{txfonts}
\usepackage{xcolor}

\journalname{Experimental Astronomy}
\begin{document}

\title{Ultra-Low-Frequency Radio Astronomy Observations from a Selenocentric Orbit}
\subtitle{First results of the Longjiang-2 experiment}


\author{Jingye~Yan        	\and
        Ji Wu               \and
        Leonid I. Gurvits   \and
        Lin Wu              \and
        Li Deng             \and
        Fei Zhao            \and
        Li Zhou             \and
        Ailan Lan           \and
        Wenjie Fan          \and
        Min Yi              \and
        Yang Yang           \and
        Zhen Yang           \and
        Mingchuan Wei          \and
        Jinsheng Guo        \and
        Shi Qiu              \and
        Fan Wu          \and
        Chaoran Hu         \and
        Xuelei Chen         \and
        Hanna Rothkaehl 	\and
        Marek Morawski    
     }


\institute{Jingye~Yan, Lin Wu, Ailan Lan, Yang Yang\at
            State Key Laboratory of Space Weather, Chinese Academy of Sciences, Beijing, China \\
            and \\
            National Space Science Center, Chinese Academy of Sciences, Beijing, China \\
            \email{yanjingye@nssc.ac.cn}  \\
            and \\
            University of Chinese Academy of Sciences, Beijing, China \\
            \and
            Ji~Wu \at
             National Space Science Center, Chinese Academy of Sciences, Beijing, China \\
            \email{wuji@nssc.ac.cn}  \\
            and \\
            University of Chinese Academy of Sciences, Beijing, China \\ 
            \and
            Leonid~I.~Gurvits \at
            Joint Institute for VLBI ERIC, Dwingeloo, The Netherlands \\
            and \\
            Delft University of Technology, Delft, The Netherlands \\
            \email{lgurvits@jive.eu} \\
            ORCID: 0000-0002-0694-2459  \\               
            \and
            Li~Deng, Fei~Zhao, Wenjie~Fan, Li~Zhou, Min~Yi, Zhen~Yang \at
            National Space Science Center, Chinese Academy of Sciences, Beijing, China \\
            \and
            Mingchuan Wei, Jinsheng Guo, Shi Qiu, Fan Wu, Chaoran Hu \at
            Harbin Institute of Technology, Harbin, China \\
            \and              
            Xuelei~Chen \at
            National Astronomical Observatories of China, Chinese Academy of Sciences, Beijing, China \\
            \and
            Hanna~Rothkaehl, Marek Morawski \at
            Space Research Centre, Polish Academy of Sciences, Warsaw, Poland \\
            \email{hrot@cbk.waw.pl}\\
            \and
         }

\date{Received: date / Accepted: date}

\maketitle

\begin{abstract}

This paper introduces the first results of observations with the Ultra-Long-Wavelength (ULW) -- Low Frequency Interferometer and Spectrometer (LFIS) on board the selenocentric satellite Longjiang-2. We present a brief description of the satellite and focus on the LFIS payload. The in-orbit commissioning confirmed a reliable operational status of the instrumentation. We also present results of a transition observation, which offers unique measurements on several novel aspects. We estimate the RFI suppression required for such a radio astronomy instrumentation at the Moon distances from Earth to be of the order of 80 dB. We analyse a method of separating Earth- and satellite-originated radio frequency interference (RFI). It is found that the RFI level at frequencies lower than a few MHz is smaller than the receiver noise floor. 
\keywords{radio astronomy --
            ultra-low frequencies --
            DSL --
            FLIS --
            Longjiang-2 --
            interferometric radiometer --
            satellite array --
            lunar orbit --
}
\end{abstract}

\section{Introduction}

Pioneering observations by Karl Jansky in 1932--1933 which marked the birth of radio astronomy were conducted at the wavelength of about 14.6 m ($\sim$20.5~MHz, \cite{Jansky-1933P, Jansky-1933N}). Over the next decades, the contemporary astrophysical agenda shifted the main interest in radio astronomy research toward shorter wavelengths. This shift was also stimulated by difficulties of long-wavelength radio astronomy studies due to the radio emission propagation effects in the Earth's ionosphere and ever growing level of human-made Radio Frequency Interference (RFI).

Despite the overall shift of major attention to frequencies above $\sim$100~MHz, several radio astronomy facilities around the world were deployed in the 1960s--1990s for observations at  meter and decameter wavelengths. One of them was the Ukrainian T-shaped Radio Telescope (UTR) deployed south of Kharkiv in the 1960s that operated at wavelengths up to $\sim$40~m \citep{Konovalenko+2016}. At about the same time, decametric radio telescopes began operations at Clark Lake, USA \citep{Erickson+1982} and at Nan\c{c}ay, France \citep{Boischot+1980}. Some observations below 30~MHz were conducted from ground facilities located near the Earth's magnetic poles, in the south, in Tasmania, and in the north, in Canada  \citep{Bridle+Purton-1968, Caswell1976, Cane+Whitham-1977, Cane-1978, Reber1994, Roger+1999}. All these studies provided useful insight into the spectrum of cosmic emission at low frequencies. But they also confirmed that detailed studies of celestial radio sources in this spectrum domain are severely affected by the ionosphere and RFI. Over the past two decades, the interest to cosmological and astrophysical phenomena in the long wavelength spectrum domain triggered construction of several large meter-wavelength radio astronomy facilities such as LOFAR (Low Frequency Array)\footnote{http://www.lofar.org, accessed 2021.03.20} with the core in the Netherlands and remote stations distributed throughout Europe from Latvia in the east to Ireland in the west, the LWA (Long Wavelength Array)\footnote{http://lwa.unm.edu, accessed 2021.03.20} in the southwestern USA, and the MWA  (Murchison Widefield Array)\footnote{http://mwatelescope.org, accessed 2021.03.20} in Western Australia. These facilities, while being prime science instruments in their own right, fulfil also the role of pathfinders for the next generation large radio astronomy facility, the SKA (Square Kilometre Array)\footnote{http://astronomers.skatelescope.org, accessed 2021.03.20}. However, despite all their advanced parameters, these facilities will not be able to conduct radio astronomy studies at ultra-long-wavelengths (ULW), longer than $\sim$10~m. The latter spectral domain necessitates a dedicated radio astronomy facility above the ionosphere at the electromagnetically quiet place, either shielded from anthropogenic RFI sources or far enough from them to minimise the RFI level. 

Several attempts were undertaken in the ULW domain using space-borne instruments in the 1960s-1970s. These involved experiments conducted with the IMP-6 satellite \citep{Brown1973}, the Radio Astronomy Explorers RAE-1 \citep{Alexander+Novaco-1974} and RAE-2 \citep{Alexander+1975}. These experiments proved the existence of strong kilometer wavelength radio emission of Earth and therefore confirmed the need of ``shielding'' from the terrestrial emission in future radio astronomy observations. In particular, the RAE-2 observations have demonstrated the efficiency of the Moon as a ``shield'' from the terrestrial radio emission \citep{Alexander+1975}. The first wave of space-borne ULW observations provided also the full-sky distribution of radio brightness at frequencies below 10~MHz \citep{Novaco+Brown-1978}.

A number of follow-up studies addressed various scientific and technological issues of the ULW astronomy in space -- see \citet{ALFIS-1992, ESA-sci1997, ALFA-1998ASPC, Jester-Falcke-2009, Mimoun+2012, Boonstra+2016, Burns+2017, Belov+2018, Chen-Aminaei+2018, Bentum+2020, Chen+2019, Koopmans+2019}. Practical implementation of the new generation of ULW experiments in space began from the launch of the Chinese Lunar exploration mission {\sl Chang'E-4} in May 2018. This mission carried three ULW experiments: the Low-Frequency Interferometer and Spectrometer (LFIS) on a pair of micro-satellites, {\sl Longjiang-1} and {\sl Longjiang-2}, placed on a selenocentric orbit; the Netherlands-China Low-Frequency Explorer (NCLE) on board the {\sl Queqiao} relay satellite, positioned at the Lagrangian point L2 of the Earth-Moon system; and the Very Low Radio Frequency spectrometer on the Lunar lander, positioned on the surface of the far side of the Moon.

The ULW LFIS experiment with the {\sl Longjiang} satellites was conceived as a pathfinder for the prospective mission DSL (Discovering the Sky at Longest Wavelengths) \cite{Ji+2015,Boonstra+2016,Chen+2019}, and other future ULW facilities. The technical objective of the experiment implemented on the two satellites was to demonstrate radio interferometry with space-space baselines, to identify strong celestial ULW radio sources, verify the impact of Earth-originated RFI and determine the global sky spectrum at the ULW domain of the spectrum. 

The {\sl Longjiang-1} and {\sl Longjiang-2} satellites described in this paper were launched by the {\sl CZ-4C} space rocket together with the relay spacecraft {\sl Queqiao} of the Chinese {\sl Chang'E-4} Lunar mission on 2018 May 21. The initial concept of the LFIS experiment aimed to demonstrate space-borne radio interferometry at frequencies below 30~MHz. Both satellites should have to fly along the same selenocentric orbit with the interferometric baseline between them re-configurable from 1~km to 10~km by low thrust maneuvers applied to accelerate or decelerate one of the satellites.

Unfortunately, due to a malfunction of the thruster control logic, {\sl Longjiang-1} was lost on the Earth--Moon transfer trajectory. Thus the LFIS experiment was conducted with only one satellite, {\sl Longjiang-2} in a ``single telescope'' mode without interferometric components of the project. The LFIS experiment was conducted during about a year on the selenocentric orbit. After completion of the scientific program, the {\sl Longjiang-2} satellite was controllably disposed onto the lunar surface on 2019 July 31. During its near one-year lifetime, the satellite conducted 81  LFIS observations near the lunar limb, over the far or near side of the Moon. Data downlink sessions were conducted at the near side segment of the same lunar orbit as the observation. Uploading mission commands for the LFIS observations, its data downlink and other commands for onboard operations were uploaded from the ground control station once a week.

In this paper we describe the ULW LFIS instrumentation of the {\sl Longjiang} satellites and present results of the radio spectrum measurements at frequencies from 1~MHz to 30~MHz conducted from one of the satellites, {\sl Longjiang-2}. The paper is organised as follows. Section~\ref{LFIS-Instr} provides the description of the LFIS instrumentation, including the satellite and the payload. Section~\ref{verif} introduces how the payload was verified before operations. Section~\ref{obs} describes an outcome of a typical observation. Section~\ref{disc} presents conclusions of the study.


\section{The LFIS experiment instrumentation}
\label{LFIS-Instr}

A pair of identical LFIS sets of payload instrumentation were placed aboard two satellites. In the following subsections we discuss the satellites' hardware with the emphasis on the instrumentation that has enabled spectrometric and polarimetric measurements in the hitherto little explored ULW domain of the electromagnetic spectrum.

 \subsection{The {\sl Longjiang-2} satellite}
 \label{Long-sat}
 
   \begin{figure}
   \centering
   \includegraphics[width=\hsize]{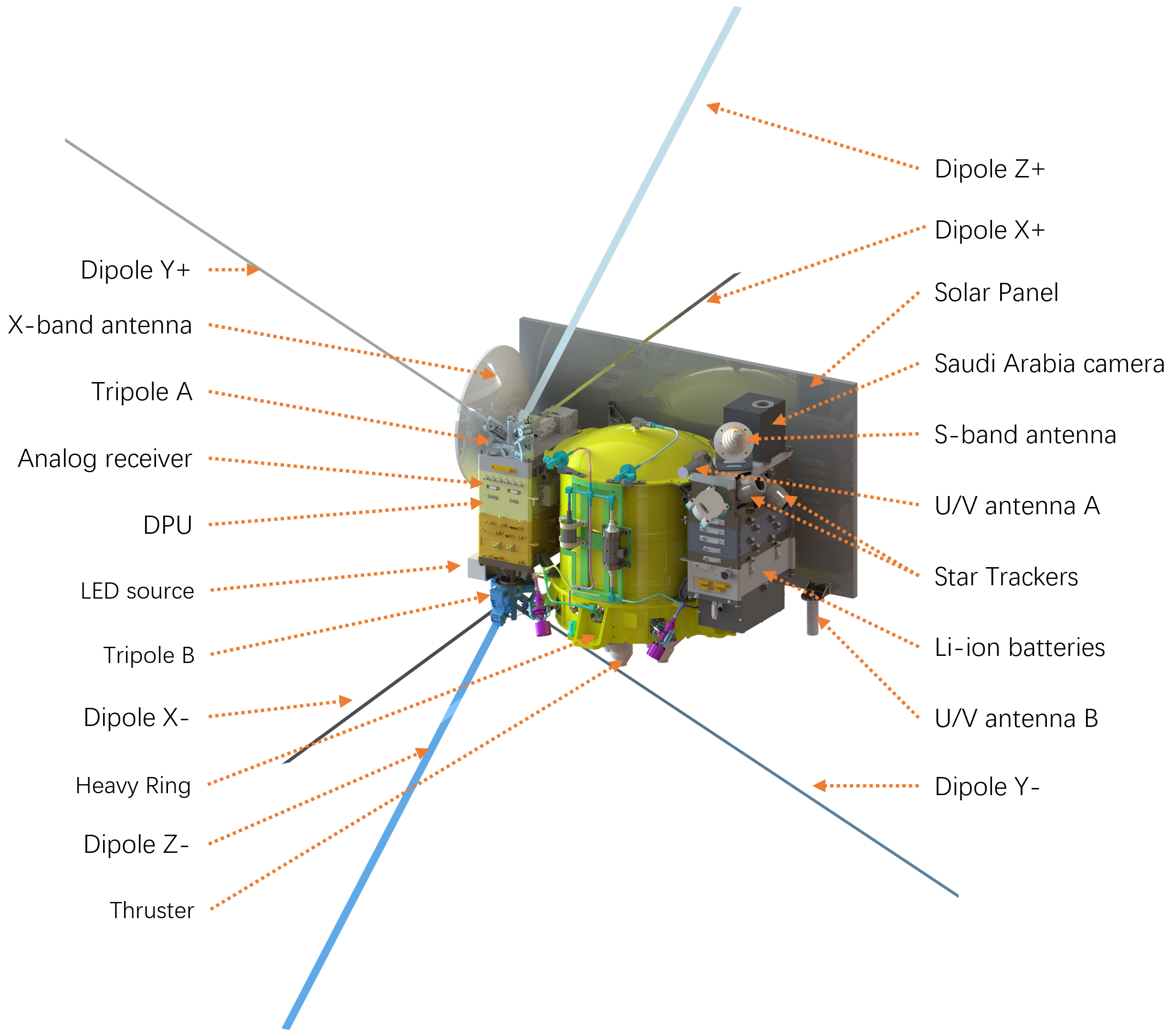}
      \caption{The {\sl Longjiang} satellite. The central cylinder of the structure is a fuel tank, which also serves as the main mechanical frame of the entire satellite. Six reaching out rods form 3 orthogonal dipoles. The length of each rod is $\sim 1$~m. Each triplet of mechanically connected rods composes a tripole antenna module. The tripole~A is on the top of the satellite (the upper side on the plot), next to the parabolic downlink X-band (8.4~GHz) antenna. The tripole~B is at the opposite (bottom) side of the structure. The LFIS electronics is placed between the vertices of the tripole antennas A and B. Boxes on the left side of the picture contain an analog receiver, DPU and the auxiliary module, which includes an inter-satellite CRS system. The Light Emission Diode (LED) source is also attached to the boxes on the left side of the structure. The boxes on the right side contain an S-band communication antenna (2.3~GHz), a Saudi Arabia optical camera, star trackers, and a VHF/UHF module for amateur radio experiments. A flat solar panel is on the back side of the structure; the satellite attitude control system keeps the latter pointed to the Sun constantly. 
              }
         \label{Longjiang-fig}
   \end{figure}
%

In order to meet the main objective of the LFIS experiment, each {\sl Longjiang} satellite carried a set of dedicated payload. It consisted of a pair of tripole ULW antennas, three front-end baluns, three analog receivers and a Digital Processing Units (DPU) as shown in Fig.~\ref{Longjiang-fig}. The antenna is composed of 6 single poles, or 3 dipoles that are orthogonal to each other. The 6 poles are mechanically arranged as a pair of tripoles called tripole-A on the top of the satellite and tripole-B attached to the opposite side of the structure as shown on the figure. To clarify the terminology used here and elsewhere in this paper, the antenna is called a {\sl tripole} in terms of its mechanical structure and a {\sl dipole} in terms of its electrical properties, its field pattern in particular. 

One front-end balun combines each pair of parallel poles together, its output feeds into an analog receiver. The DPU firstly samples each signal from 3 receivers into digital signal, each corresponding to a specific linear polarization. Then a Fast Fourier Transform (FFT) module processes 3 digital signals in parallel and provides continuum spectrum over full bandwidth, in addition to forming 10 selected narrow bandwidth signal for each polarization. These digital operations are performed with two hardware cores inside the same FPGA. Such the arrangement for the spectrum channelization was adopted in order to achieve the original aim of the mission -- cross-correlation of signals obtained with two satellites in a single-baseline interferometer. However, as described in this paper, this setup proved to be applicable for single-telescope observations too.

In addition to 3 receiving chains (one for each dipole), the instrumentation also contains an auxiliary module which carries an inter-satellite communication, ranging and synchronisation (CRS) unit. The latter's aim is to conduct signal transmission from {\sl Longjiang-2} to {\sl Longjiang-1} and clock synchronisation for coherent signal sampling. It also makes range measurements between the satellites. The baseline attitude is measured with a Light Emission Diode (LED) array placed on {\sl Longjiang-2} which creates an artificial optical source detected by the the sky camera on {\sl Longjiang-1} with the background of stars. A combination of the range and attitude measurements provides determination of the interferometer's baseline vector. Since only {\sl Longjiang-2} has reached the nominal selenocentric orbit, the auxiliary module in support to the interferometric mode of operation will not be discussed further in this paper.

{\sl Longjiang-2} inserted itself into an initial lunar elliptical orbit of $357 \times 13,704$~km on 25th May, and became the smallest spacecraft which entered lunar orbit with its own propulsion system \citep{Xibin+790}. After the insertion, a few fine-tuning maneuvers were conducted over the period of two months. 

The {\sl Longjiang} satellite platform was designed and built by the Harbin Institute of Technology (HIT), Harbin, China \citep{Xibin+790}.

\subsection{The LFIS payload}
\label{LFIS-pld}

 The LFIS payload for ULW astronomy was identical on both {\sl Longjiang} satellites.Its specification is presented in Table~\ref{tab:1}.
 
\begin{table}[h]
\caption{Main parameters of the LFIS radio system}
\label{tab:1}       
\begin{tabular}{ll}
\hline\noalign{\smallskip}
LFIS component of parameter & Specification or value \\
\noalign{\smallskip}\hline\noalign{\smallskip}
Antenna type & 2 tripoles (3 dipoles)  \\
Antenna single pole length & 1~m  \\
Antenna tip-to-tip dipole length & 2.25~m, incl. 0.25~m gap between the poles \\
Receiver & 3 channels\\
Instantaneous frequency range & 1--30~MHz \\
Dynamic range & 60~dB \\
Spectral resolution & 39~kHz \\
Filter bank & $10~\times$ 1~kHz \\
\noalign{\smallskip}\hline
\end{tabular}
\end{table}
 
   \begin{figure}
   \centering
   \includegraphics[width=0.60\textwidth]{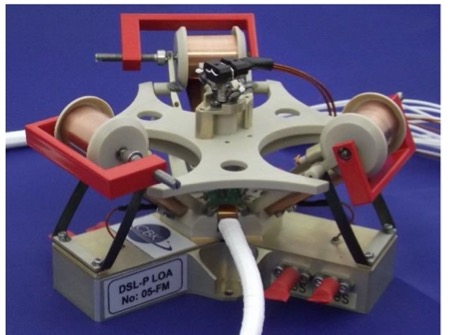}
      \caption{The antenna flight model in a folded configuration.}
         \label{pic:RELEC}
   \end{figure}
%

 \subsubsection{The LFIS antenna}
 \label{Tri-anten}
 
 Ideally, three LFIS antenna dipoles should be orthogonal to each other, and each dipole should consist of a pair of rods aligned along a straight line. However, it was difficult to implement such an ideal configuration aboard the {\sl Longjiang} satellite because of various constraints of the satellite's structure and geometry. 
 
 The LFIS antenna is based on the tripole concept developed by the Space Research Centre, Polish Academy of Sciences, for the Radio Frequency Analyser (RFA) instrument for the RELEC (Relativistic Electrons) instrument suit of the {\sl Vernov} mission \citep{RELEC-2016} led by the D.V. Skobeltsyn Institute of Nuclear Physics of M.V. Lomonosov Moscow State University (Russia). Three miniature $\varnothing$6~mm and 1~m tubular booms of the antenna are stored on respective reels and extend by spring energy released during the transformation of the flat input profile to the original round ones (Fig.\ref{pic:RELEC}).

\begin{figure}
\centering
\subfigure[Model for antenna simulations]
{
    \begin{minipage}[b]{0.4\textwidth}
        \flushleft 
        \includegraphics[width=\hsize]{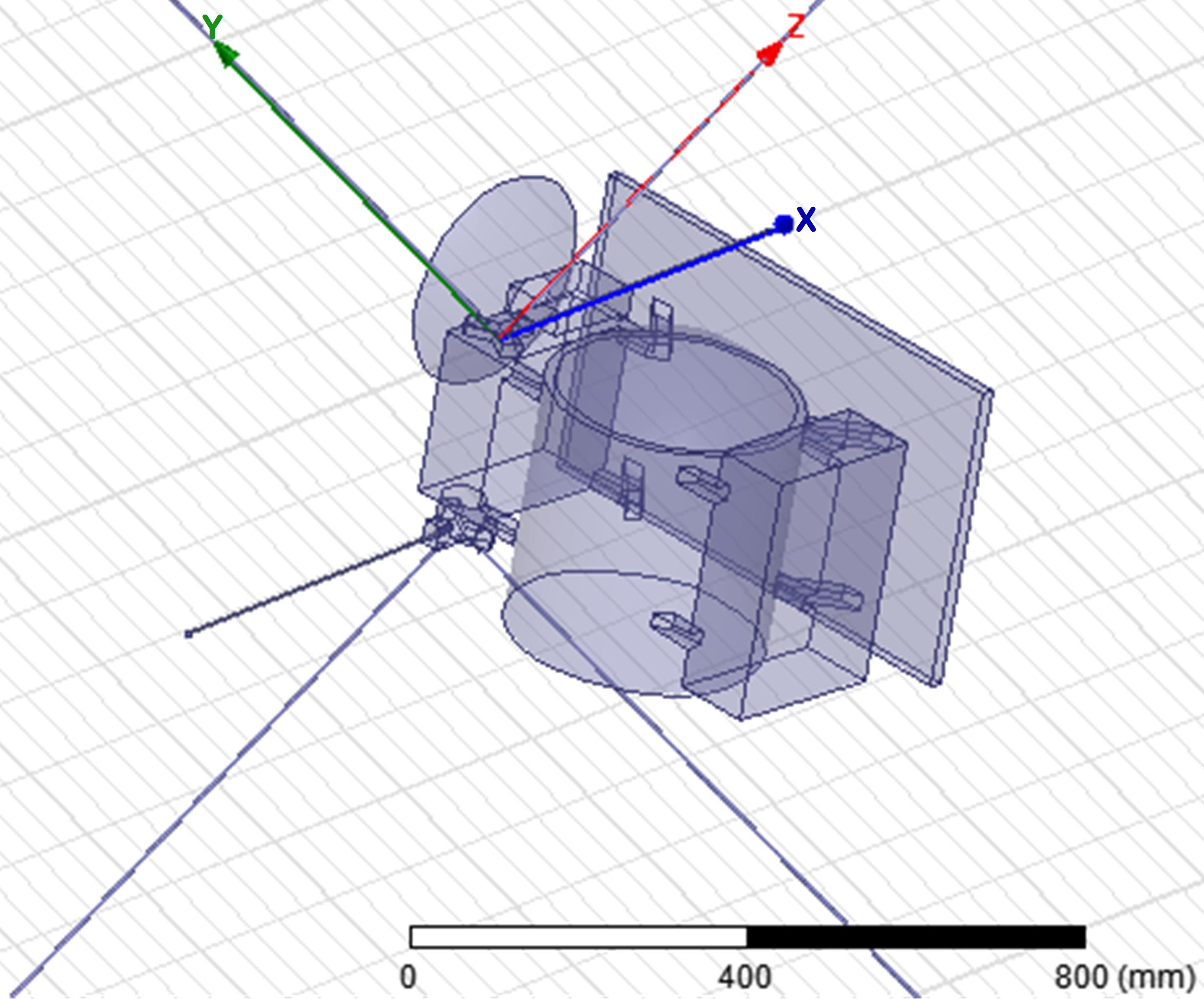}
    \end{minipage}
    \label{antenna-model}

}
\subfigure[Simulated pattern of a dipole.]
{
    \begin{minipage}[b]{0.5\textwidth}
        \flushright
        \includegraphics[width=\hsize]{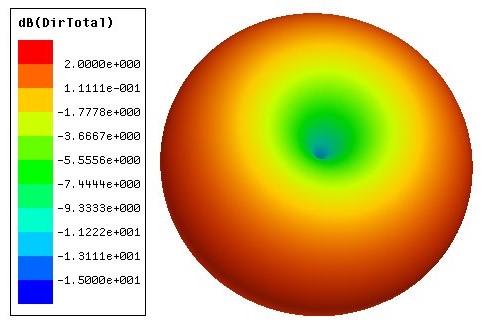}
    \end{minipage}
    \label{antenna-pattern}
}
\caption{The LFIS antenna model and pattern. (a) The simulation model incorporates three dipoles and the platform structure. (b) A simulated dipole pattern at 10~MHz that corresponds to the model (a).}
\end{figure}

 The Tripole-A and -B were developed jointly by the CBK/PAN and NSSC/CAS, and attached to the {\sl Longjiang} platform as shown in Fig.~\ref{Longjiang-fig}. A model of the tripole antennas assembled with the satellite platform is given in Fig.~\ref{antenna-model}, and a simulated antenna pattern of a dipole is given in Fig.~\ref{antenna-pattern}. The presented pattern is simulated at a frequency of 10~MHz. Although the dipoles are noncollinear, and the solar panel, downlink reflector and other components of the satellite structure in principle contribute into the interaction of the electromagnetic waves with the LFIS antennas, the simulated far field response is still perfectly symmetric, and the antenna has an omnidirectional pattern as expected. This consistency might be due to the fact that the 1~m poles are small compared even to the shortest operational wavelength of 10~m (30~MHz), and the size of the satellite (\textless ~0.5~m) in cross-section is also small compared to the length of the dipole, $\sim 2.25$~m. Thus, the structural imperfection of the tripole arrangements on the {\sl Longjiang} satellites has little effect on the far field electrical response. 
 
Since the dipole~Y was close to the plume area of the main thruster (see Fig.~\ref{Longjiang-fig}), it was desirable to deploy the Y- after completion of orbital maneuvers (sub-section \ref{Long-sat}). Thus, the tripole-B was designed with a special mechanism operating in two deployment phases. In phase~1, the poles which do not interfere with the plume, i.e., X- and Z-, have been deployed immediately after the {\sl Longjiang-2} insertion into its selenocentric orbit, and the payload was powered on for conducting in-orbit commissioning. In phase~2, Y- was deployed and the tripole-B was  fully functional after the completion of all orbital maneuvers. After phase~2, the payload was in a fully operational condition.
 
 \subsubsection{Analog receiver}
 \label{ana-rx}
 
 The analog receiving chain of LFIS is composed of six front amplifiers, one for each pole, three baluns, each combining a pair of parallel poles and forming three orthogonal signals, followed by three back-end amplifiers which prepare the signal for the digitizer. Such a rather unusual arrangement was dictated by the constraints of the antenna lay-out that required the dipoles' arms to be separated from each other.
 
   \begin{figure}[h]
   \centering
   \includegraphics[width=0.60\textwidth]{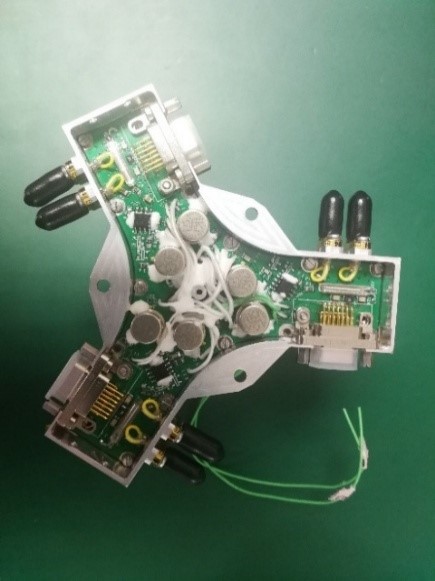}
      \caption{The front-end of a tripole. The Y-shaped structure contains 3 independent LNAs, each occupying one arm. Two feeds per arm are used for the analog signal output and calibration signal input, respectively.
              }
         \label{front-end}
   \end{figure}
%
 
\begin{figure}
   \centering
   \includegraphics[width=\hsize]{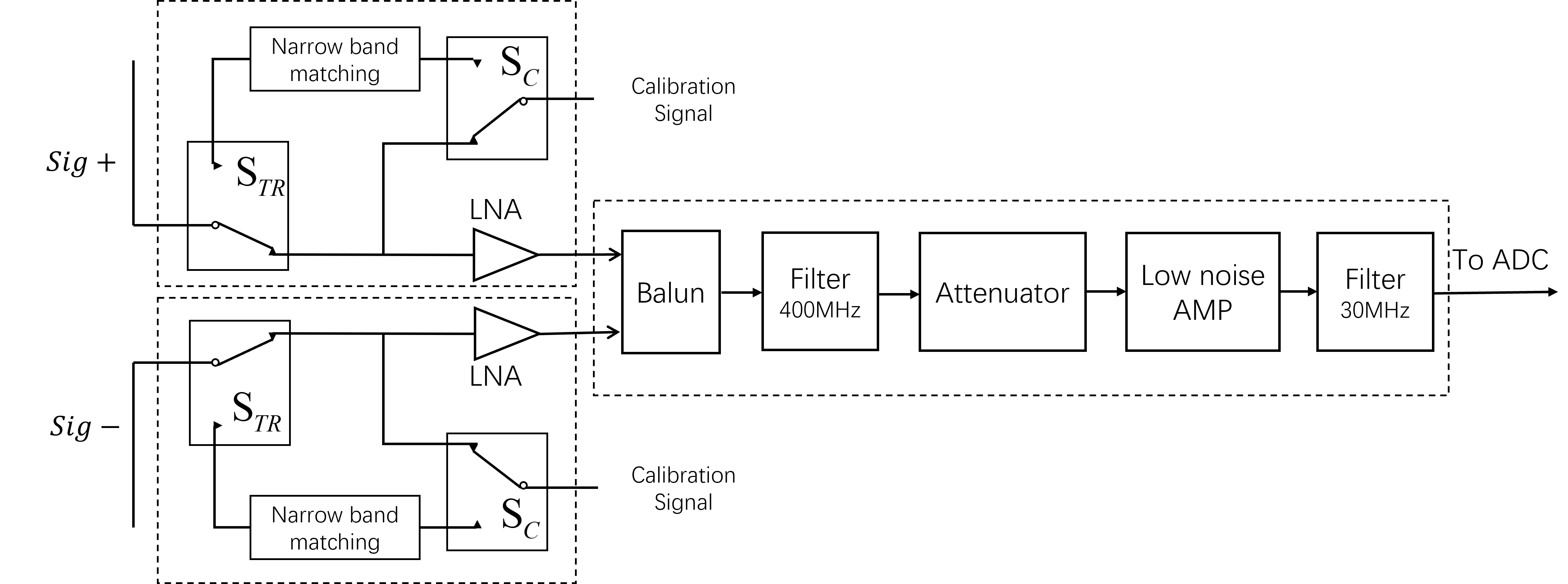}
      \caption{The block-diagram of an analog receiver chain. The left two dashed squares represent a pair of amplifiers for a pair of poles that are aligned as a dipole. The right dashed square represents the signal combination by a balun followed by an amplifier and an analog low pass filter. The output of the receiver, together with the ones from the other two receivers, is connected to the DPU.
              }
         \label{receiver-fig}
   \end{figure}

The front-end module is designed for ultra wide bandwidth matching (1--30~MHz) with a low noise FET-input operational amplifier OPA657, which is capable of amplifying the signal with high impedance of $\sim$1~M$\Omega$ in front of the amplifier. It is unusual for radio astronomy to use high impedance input because of its high noise. However, it is acceptable for the ULW domain, since the sky brightness temperature dominates the system temperature\footnote{A detailed analysis of the system temperature and the sky temperature measured by LFIS is outside of the scope of this paper and will be addressed elsewhere. However, we note that our preliminary estimates indicated the LFIS antenna temperature was below $10^4$~K, while the sky temperature at 30~MHz was about $10^4$~K smoothly rising to $7\times 10^7$~K at 6~MHz.}. Each tripole is connected to a common Y-shaped enclosure which hosts three front-ends and calibration components (Fig.~\ref{front-end}). There are two connectors applied to each arm of the cavity, one for sky signal output to the balun, the other for reference signal input to the front-end for stability and similarity calibration. The block-diagram of the front-end is given in the left part of the scheme surrounded by a dashed line on Fig.~\ref{receiver-fig}. In the radiometry mode, the sky signal goes through the switch (shown as $S_{TR}$ in Fig.~\ref{receiver-fig}) between the narrow band matching and Low Noise Amplifier (LNA), and further into the receiver. In the internal calibration mode, the calibration signal generated by a Digital to Analog Converter (DAC), is directed into the LNA through the calibration switch $S_C$. In the external calibration mode, the calibration signal is directed through $S_T{}_R$ and $S_C$ to the antenna and transmitted to the other satellite ({\sl Longjiang-1} in the discussed case). The narrow band matching network between two switches is applied in order to provide an acceptable standing-wave ratio and maximize the radiated power within a narrow band about 28~MHz. Due to the loss of the {\sl Longjiang-1}, it was not possible to demonstrate the external calibration procedure in real space flight conditions.

%

 Six outputs from the front amplifiers, coming from 2 tripoles, are connected to a common back-end amplifier module. Each pair of signals from a dipole is combined by a balun and further amplified to a proper voltage for the digitizer. A 400~MHz low pass filter positioned between the balun and LNA is used to suppress high frequency interference generated by S- and X-band communication systems of the satellite. An independent low pass filter with stop–band at 30~MHz positioned after the LNA shapes the frequency response of the receiver chain. Such the two-filter configuration enables efficient RFI suppression potentially caused by the dual-band radio communication system. An adjustable attenuator is included in each chain of the back-amplifier module in order to prevent the receiver saturation and suppress any unexpected spikes, which might have a natural cause (a very strong received signal) or an impact of RFI. The attenuator is configured by a command generated on the ground. Fast switching of the attenuator might result in the so called 'ripples' in the signal parameters. Therefore, measurements are made after the the attenuator relaxes into a stable state. Output of the analog receiver is digitized by high speed ADC operating at 80~Msps for further digital processing.

\subsubsection{Digital signal processing}
   \begin{figure}
   \centering
   \includegraphics[width=\hsize]{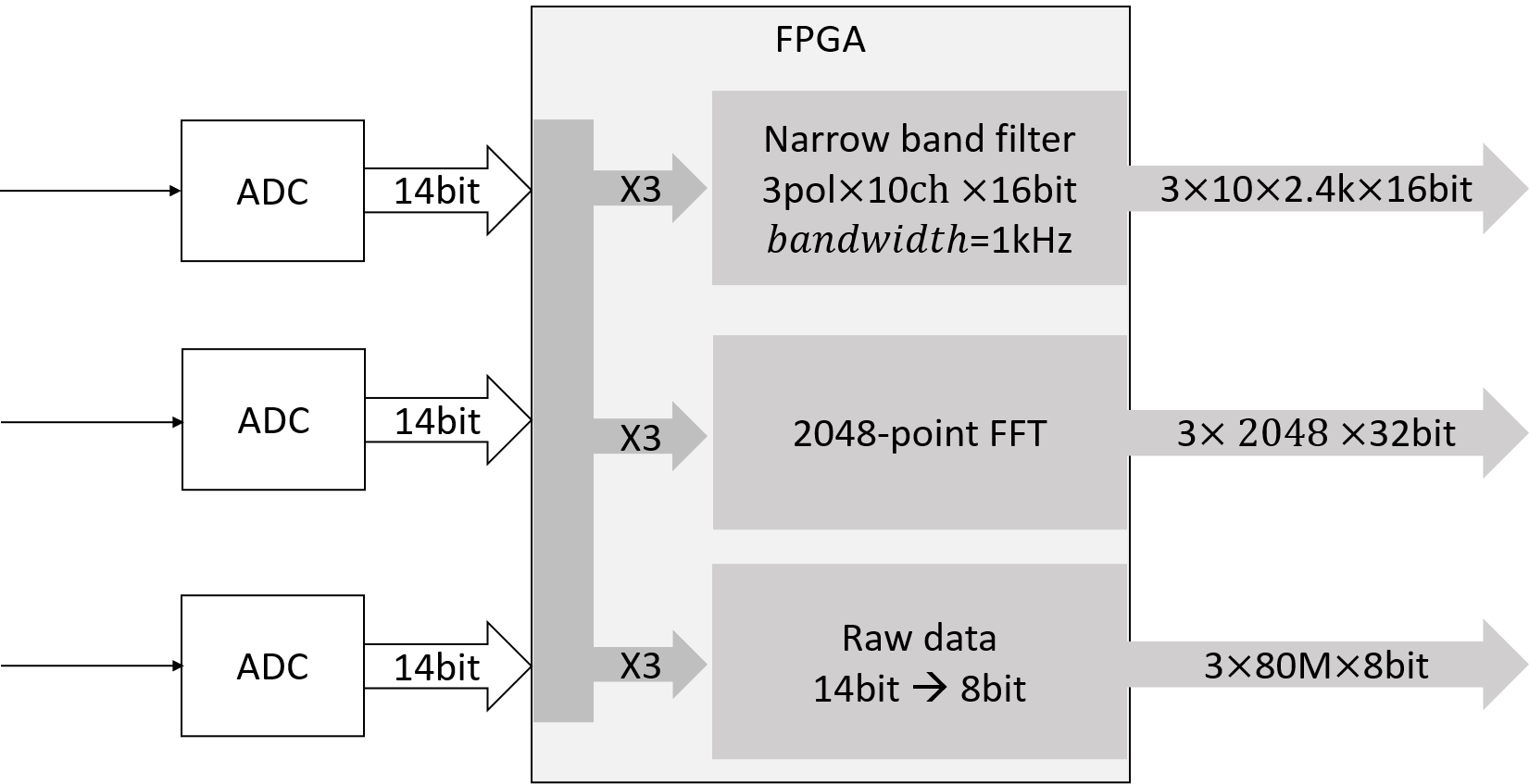}
      \caption{The on-board DPU includes a high speed data bus which feeds the digitized samples into the filter-bank and FFT core in parallel. Under commands of the ground-based control, the data bus can also be connected to the raw Data core in the same FPGA for storing the sampling stream into the massive memory of the satellite.
              }
         \label{signal-handling}
   \end{figure}
%

The LFIS ADC operates with 14-bit encoding. However, since the noise and spurs generated by the processing board contaminate lower bits, only 12 most significant bits are considered to be reliable. Digital samples of three signals are handled in parallel with two processing cores implemented in the same FPGA.

The first core is a filter-bank with 10 narrow-band channels of 1~kHz bandwidth. All narrow-band data were downloaded to the ground station. It was supposed that the data from two satellites would be cross-correlated on the ground. At the same time, a single tone signal generated by the calibration DAC will periodically scan over 10 channels by adjusting the output frequency sequentially, receiver outputs of the injected single tone were also downloaded for amplitude and phase calibration. Bandwidth of 1~kHz is limited by the so-called fringe-washing effect (see \cite{MIRAS2003} for details), 
which will cause bandwidth smearing. The number of the filter channels is defined by the capacity of the data down-link system, 1~Mbps. This core deals with signals from three dipoles, each served by 10 narrow-band channels. The filtered data are represented by 16~bit streams with the data rate of 2.4~k samples per second per channel. The data rate produced by this core is therefore $3\times10\times2.4\times16 =$ 1.152~Mbps. 

The second core is responsible for calculating a continuum spectrum with the resolution of 39~kHz permitted by the sampling rate of 80~Msps. This is achieved by a Fast Fourier transform (FFT) algorithm with 2048 data points and natural windowing. More efficient but more sophisticated windows were not used due to the strict power budget constraints onboard the satellite. This deemed to be suitable for the mission aiming to detect strong sources with little or no spectral variations. The capability of this core permits only a 27.7\% duty cycle, or 0.31~s of spectral integration every 1.12~s cycle period. This results in disruptions of the data processing limit the temporal resolution to 1.12~s. The FFT core produces the data with the rate $3\times2048\times32 \approx$ 196k bits in 0.31~s, with the average rate about 54~kbps. 

During the in-orbit commissioning phase, a special mode of data handling, the so called raw data mode, was employed. In this mode, the signal was sampled with the rate of 80~Msps for a short period and all the samples were downloaded. The raw data are useful for comprehensive diagnostic purposes. Only a small set of data covering less than 1~s of observing time has been down-linked to the ground since its massive data volume. Analysis of this data set confirmed that the instrumentation was functioning in-flight in the same way as in pre-launch ground tests. In order to reduce the amount of down–linked data, the test digital stream sampled originally with 14 bits was truncated to 8~bits. The aggregate data rate in this mode was therefore $3\times 80\times 8 = 1.92$~Gbps.

To summarize, the LFIS data were provided in 10 narrow-band channels sampled continuously during an observation. The signal spectrum was obtained with the integration time 0.31~s every 1.12~s. Fig.~\ref{signal-handling} illustrates the signal processing data flow.

\section{In-orbit verification}
\label{verif}

The {\sl Longjiang-2} satellite was inserted in the selenocentric orbit on 2018 May 26. The initial orbital period was 20.44~h. The payload was powered-on on 2018 May 29, the tripole-A was deployed immediately after that. The Tripole-B was thoroughly deployed on 2018 September 14 after completion of the orbital manoeuvre (as explained in sub-section \ref{Tri-anten}). At that point, the payload was declared fully operational. 

\begin{figure}[t]
   \centering
   \includegraphics[width=\hsize]{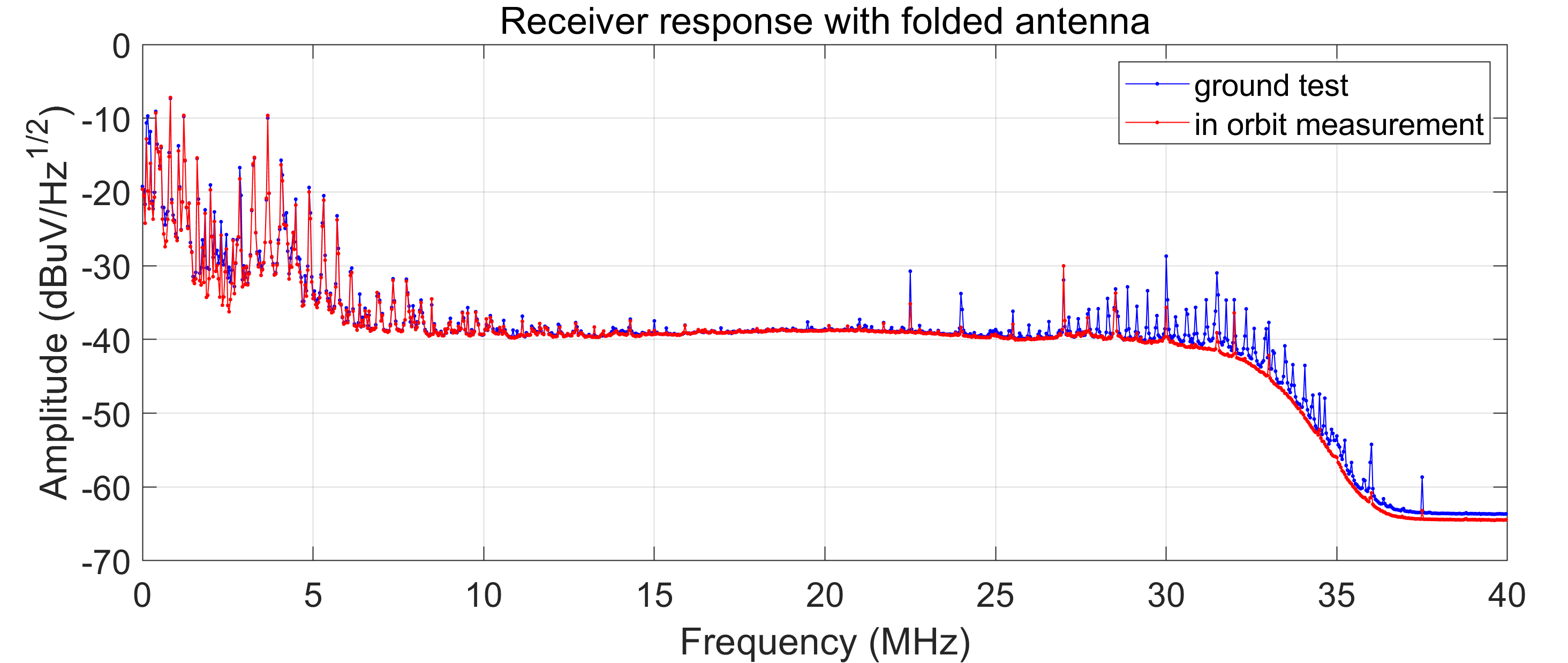}
      \caption{The spectrum measured inside the Maxwell chamber at the ground-based facility and in orbit. The antenna is folded in both cases, therefore the spectrum represents a combination of the instrument noise floor and local RFI contribution, turns up below ~6 MHz due to frequency leakage, from satellites electronics. The vertical axis scaled in physical units at the antenna feed.  
              }
         \label{receiver-response}
   \end{figure}

The first examination of the payload was conducted before the antenna deployment. This allowed us to obtain the spectrum formed by the receiver own noise and its response to the electromagnetic interference (EMI) generated by the satellite's systems. This test resulted in almost identical spectra from all three receivers. As an example, we show one of them in Fig.~\ref{receiver-response}. Compared to the pre-launch measurements obtained in a Maxwell chamber, with anechoic absorber that does not work well at low frequencies of our interest, the data from the in-orbit verification test confirmed a good status of the instrument. Both the ground and in-orbit measurements demonstrate a similar noise floor, EMIs are still concentrated at the opposite lower and higher frequency ends of the spectrum. In fact, the latter noise pattern in the in-orbit measurements appeared somewhat better than the similar one in the ground test, especially at the frequencies around 30~MHz. It suggests that the EMI environment in the ground facility and in space are slightly different. The spectrum drops off at the higher frequencies as the response to the filter. A very strong RFI at the lower frequency comes from the satellite's electronics. Obviously, at the lower end below 6~MHz, the spectrum seems turning up. This is actually an artifact due to the frequency leakage of the FFT algorithm, which spread the strong interference into neighboring spectrum channels. This conclusion is demonstrated in Fig.~\ref{narrowband}, where 10 ultra-fine spectra of 1~kHz bandwidth at different central frequencies present approximately the same amplitude. Since the ultra-fine spectra are acquired with digital filters of high stop-band suppression, they suffer little frequency leakage. 

 These first in-flight measurements had demonstrated that the entire receiver chain was in nominal operational conditions after the insertion into the selenocentric orbit.  

%

\section{Observations}
\label{obs}

The power budget and downlink capacity limitations of the {\sl Longjiang} micro-satellite made possible ULW observation lasting $\sim$10 to 20~min every selenocentric orbit. Most observations were conducted either starting at the {\sl Longjiang's} visible apparent position near the Moon's limb and ending shortly behind the Moon, or in the reverse order. The limb is a natural transition border of the Moon's RFI shielding. 

\begin{figure}
\centering
\subfigure[3D trajectory evolution of {\sl Longjiang-2} over 10 days.]
{
    \begin{minipage}[b]{0.9\textwidth}
        \flushleft 
        \includegraphics[width=\hsize]{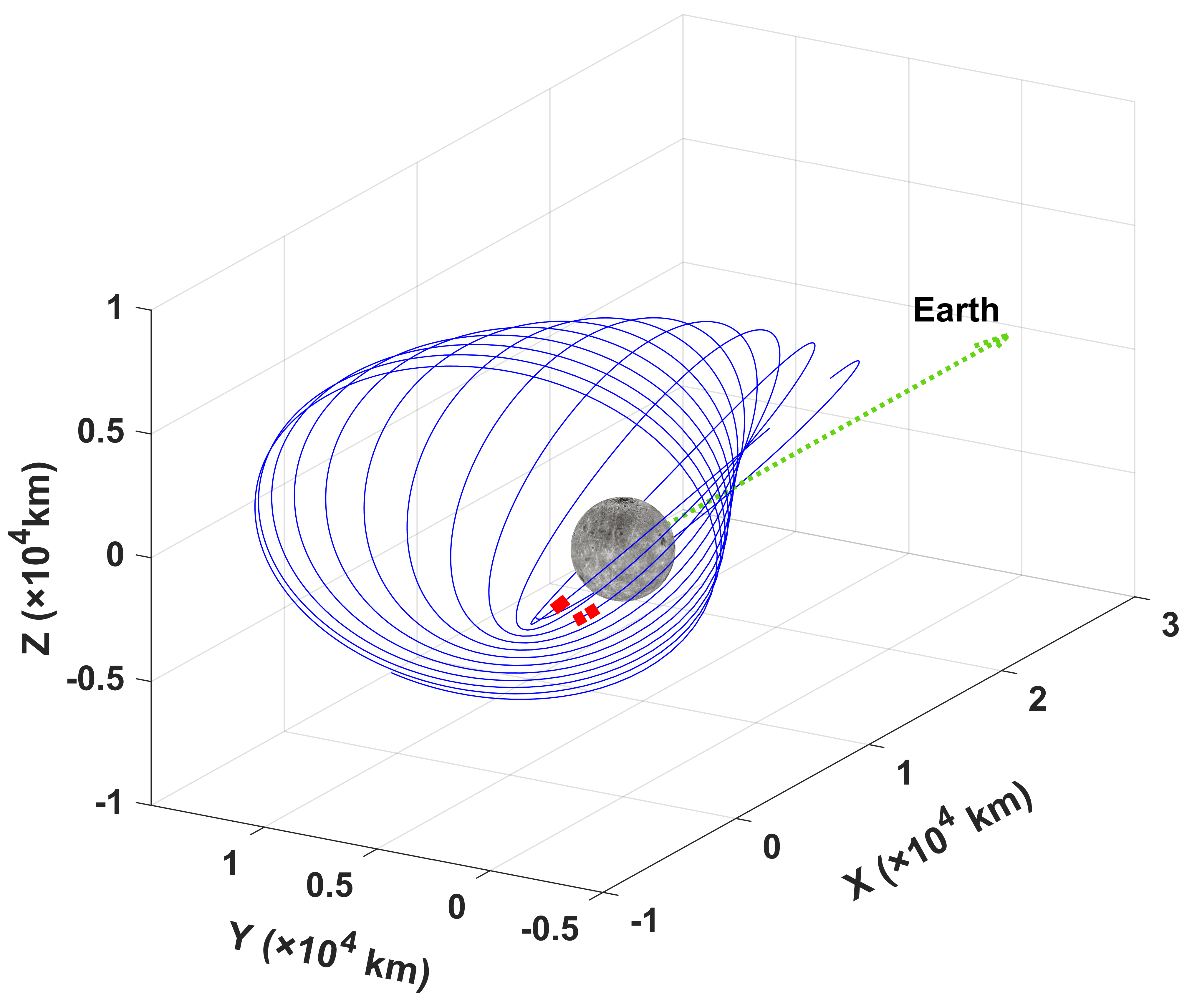}
    \end{minipage}
    \label{3D-orbit}
} 
\subfigure[Projection of the trajectory of {\sl Longjiang-2} on the $x$-$y$-plane over 10 days.]
{
    \begin{minipage}[b]{0.9\textwidth}
        \flushright
        \includegraphics[width=\hsize]{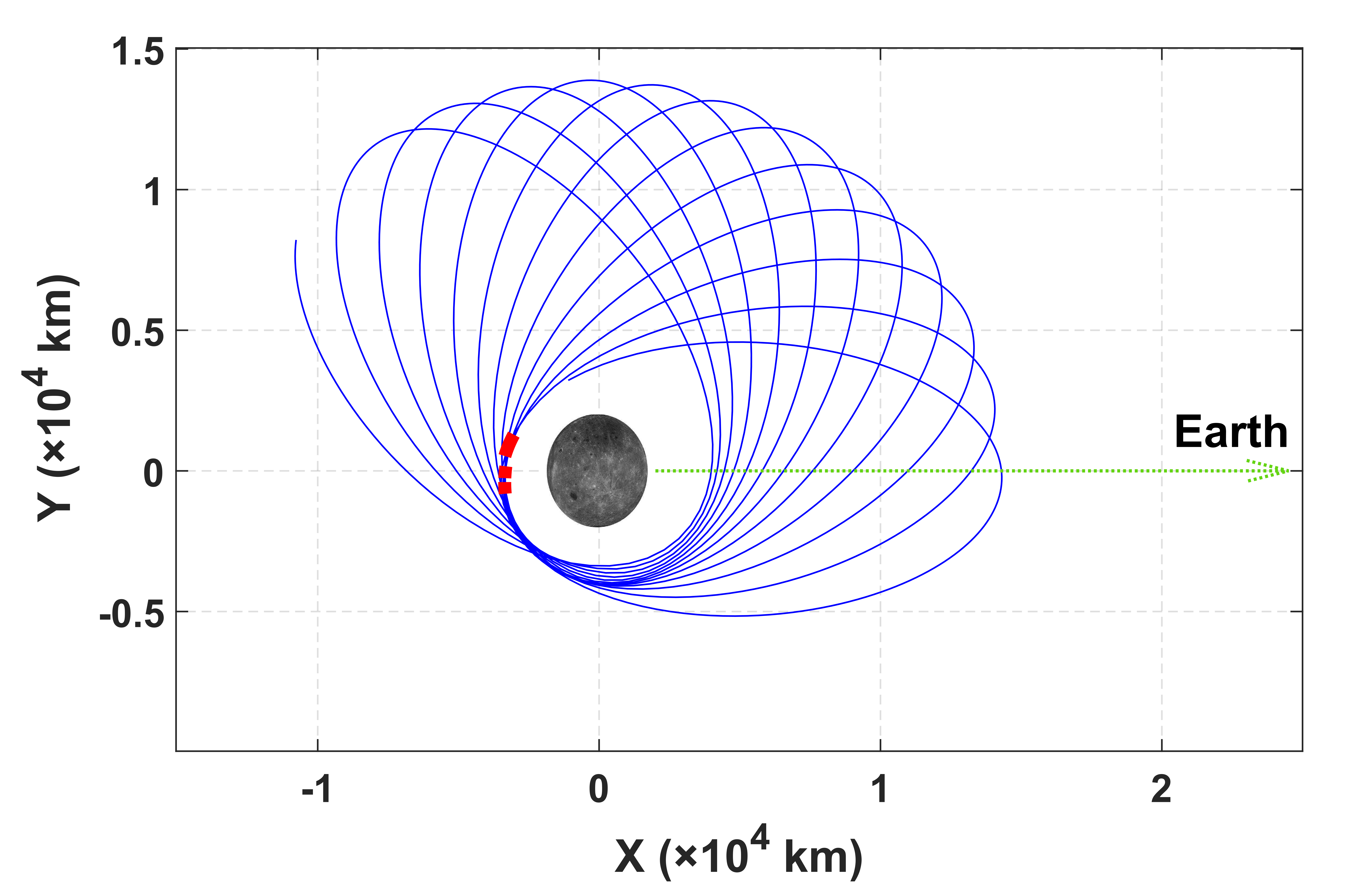}
    \end{minipage}
    \label{2D-Jul}
}
\caption{The selenocentric orbit of {\sl Longjiang-2}. In the adopted coordinate system, the line connecting the Moon's and Earth's centers of mass is the $x$-axis. The tangent line of the geocentric Moon's orbit is defined as the $y$-axis. The $z$-axis is normal to the plane defined by the axes $x$ and $y$ and points toward the northern hemisphere. The direction of the $y$ axis is chosen to make the system right-handed. The origin of the system is located at the Moon's center of mass. The {\sl Longjiang-2} trajectory is shown in blue. The red arcs indicate the position of the satellite during observations. (a) The 3D orbital evolution over about 10~days. (b) The {\sl Longjiang-2} trajectory (shown in blue) projected on the $x$-$y$-plane between 27 July and 7 August, 2018. The red segments indicate the position of LFIS measurements.
}
\end{figure}

The geometry of the selenocentric trajectory of {\sl Longjiang} is shown in Fig.~\ref{3D-orbit} in a coordinate system described in the figure caption. 
In these arbitrary coordinates, $x$-axis constantly points to the Earth in order to explicitly show the Moon's shielding, thus the elliptical orbit rotates around the origin (practically -- the center of mass of the Moon) following the periodic movement of the Moon one cycle per lunar month. 

The orbit precession effect illustrated in Fig.~\ref{3D-orbit} helps to rotate the orbital plane around the $z$-axis, thus gradually the interferometer's $(u,v,w)$-space becomes filled densely and isotropically. The observations are indicated as red arcs on the trajectory. These arcs are very short compared to the total selenocentric orbital period of 20.44~hrs. The observations are mostly taking place around the Moon's limb at different distances, except a few taken without the Moon shielding. Fig.~\ref{2D-Jul} shows the projected trajectory onto the $x-y$~plane within 10 days. 

\begin{figure}[h]
   \centering
   \includegraphics[width=\hsize]{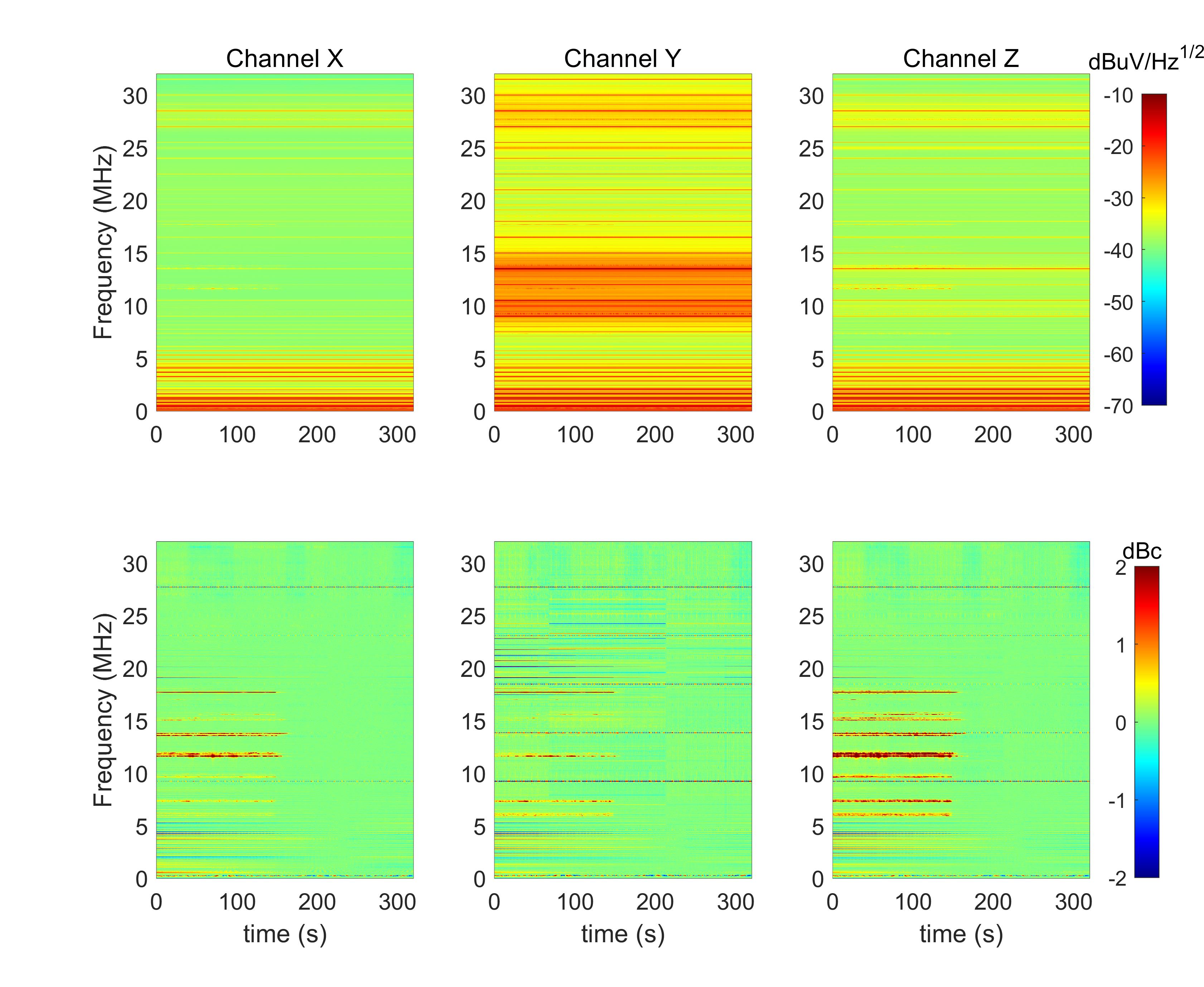}
      \caption{Measured spectra during the first occultation transition. The top 3 panels show the measured spectrum with 3 orthogonal dipoles. Definitions of X, Y and Z can be found in Fig.~\ref{Longjiang-fig}. The 3 bottom panels show the corresponding spectra after subtracting the average of the occulted period, i.e., the measurements from ~160~s to 300~s. The subtraction removes the satellite interference as well as the global celestial signal. The residual signal is the terrestrial RFI. The top three panels represent received voltage at he feed of each dipole. The lower three panels represent the relative power. The terrestrial interference is about 2~dB higher than the global spectrum as seen in the bottom panels. 
              }
         \label{spectrum}
   \end{figure}

The very first occultation transition was measured on 2018 September 28 and is presented in Fig.~\ref{spectrum}. In this plot, the transition takes place at about 160~s. A significant electromagnetic interference is present after the antenna deployment. Strong spikes still appear at frequencies below ~6~MHz and around 30~MHz. These are consistent with the measurements conducted during the commissioning with the folded antenna as illustrated in Fig.~\ref{receiver-response}. This means the satellite-originated RFI dominates over the terrestrial interference. The measurements in channels X and Z are similar to those with the folded dipole, while channel Y is much more contaminated after the deployment. It is natural to suggest that the interference received in channel X is dominated by the conductive interference therefore less impacted by the deployment, and channel Y is severely influenced by spatial coupling through the deployed dipole. The channel Z demonstrates the level of interference between those of channels X and Y. Different interference characteristics of the three channels provide evidence of the effects of spatial coupling. It might be a result of the satellite's structural configuration, including the solar panel and downlink antenna located at different positions with respect to the three LFIS dipoles, as well as the impact of the onboard electronics. Although the LFIS dipoles' far field pattern is little impacted, the near field spatial coupling of the local RFI is obvious.

%

The spectra shown in three bottom panels of Fig.~\ref{spectrum} are derived by subtracting the average signal of the occulted period from the single 300~s-long observation. This manipulation results in removal of the satellite-originated interference and the received global signal from the raw data shown in the top panels. Since the receiver noise level is about 1~megakelvin (MK), much smaller than the celestial signal, we ignore the receiver noise in following analysis. If the local satellite-originated interference and the celestial signal are constant, only the Earth-originated interference remains after the subtraction. However, at least 3 persisting  RFI lines at 9, 14 and 28~MHz are still evident as shown in the bottom panels, especially in Channel Z. The RFI in these lines appears to be temporal varied satellite-originated RFI. The three channels demonstrate different level of RFI after the subtraction, with channel X cleaner than channel Z, and channel Y worse among all three. The interference below ~6~MHz shows a long term variation without the cut-off signature by shielding.  

The limb transition takes place at about 160~s as is evident in the bottom panels of Fig.~\ref{spectrum}.
These spectra also indicate that strong terrestrial RFI is mainly concentrated between 6 and 20~MHz for this observation. The bright RFI lines are roughly a few dB higher than the celestial signal. After the subtraction, the noise floor below 6~MHz is similar before and after the shielding transition, i.e., no visible Earth-originated RFI is found. It suggests that, at very low frequency, the majority of Earth-originated EMIs are sufficiently blocked by Earth's ionosphere. The bottom panels also show slightly different transition time at different frequencies. This is reasonable due to the fact that the RFI sources are separately distributed over the Earth surface. The power of terrestrial RFI almost immediately cuts off below the noise floor after crossing the limb. However, further studies with instrumentation providing higher spectral and temporal resolution is required for detection and studies of the terrestrial RFI diffraction and dispersion. 

In the first observation, the terrestrial RFI is about 2~dB higher than the system temperature (the sky signal temperature plus receiver temperature). Assuming the sky temperature of about 1~MK at the observing low frequencies, and the RFI brightness of comparable temperature, the theoretically predicted cosmological spectrum variation of Dark Ages at the level of tens of millikelvin is 80~dB lower than the RFI signal.  
The spectra in Fig.~\ref{spectrum} also suggest that for a lunar orbiter, as long as the interference is temporally stable, 
terrestrial RFI can be mitigated efficiently by subtracting it from the raw data using the signal containing local RFI detected during the occulated period. 
Moreover, for the same terrestrial RFI, the relative power level between dipoles is different. For example, around 12~MHz, the detected interference in channel Z is larger than that in channel X, and the latter is larger than that in channel Y. Meanwhile, around 8~MHz, the interference in channel Z is larger than that in channel Y, and the latter is larger than that in channel X. The reason of this discrepancy might be purely geometrical (e.g., a difference in the effective length of dipoles) or indicate that the RFI signal is polarized. Future RFI mitigation techniques will have to involve accurate enough calibration enabling robust distinction between these two options. 

\begin{figure}
\centering
\subfigure[Egress on 2018 Oct. 01.]
{
    \begin{minipage}[b]{0.8\textwidth}
        \flushleft 
        \includegraphics[width=\hsize]{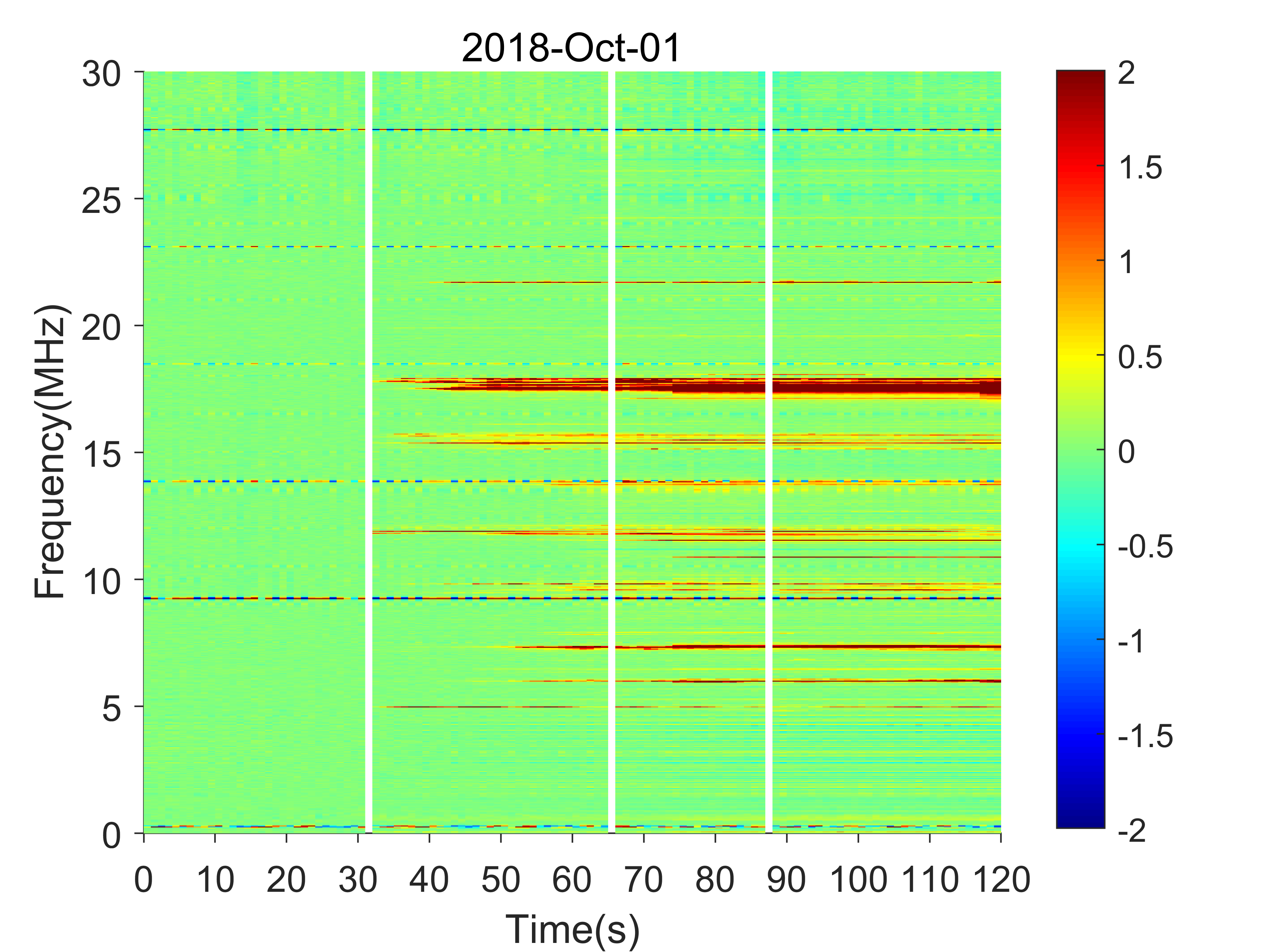}
    \end{minipage}
    \label{data-1001}
}

\subfigure[Ingress on 2018 Oct. 03.]
{
    \begin{minipage}[b]{0.8\textwidth}
        \flushright
        \includegraphics[width=\hsize]{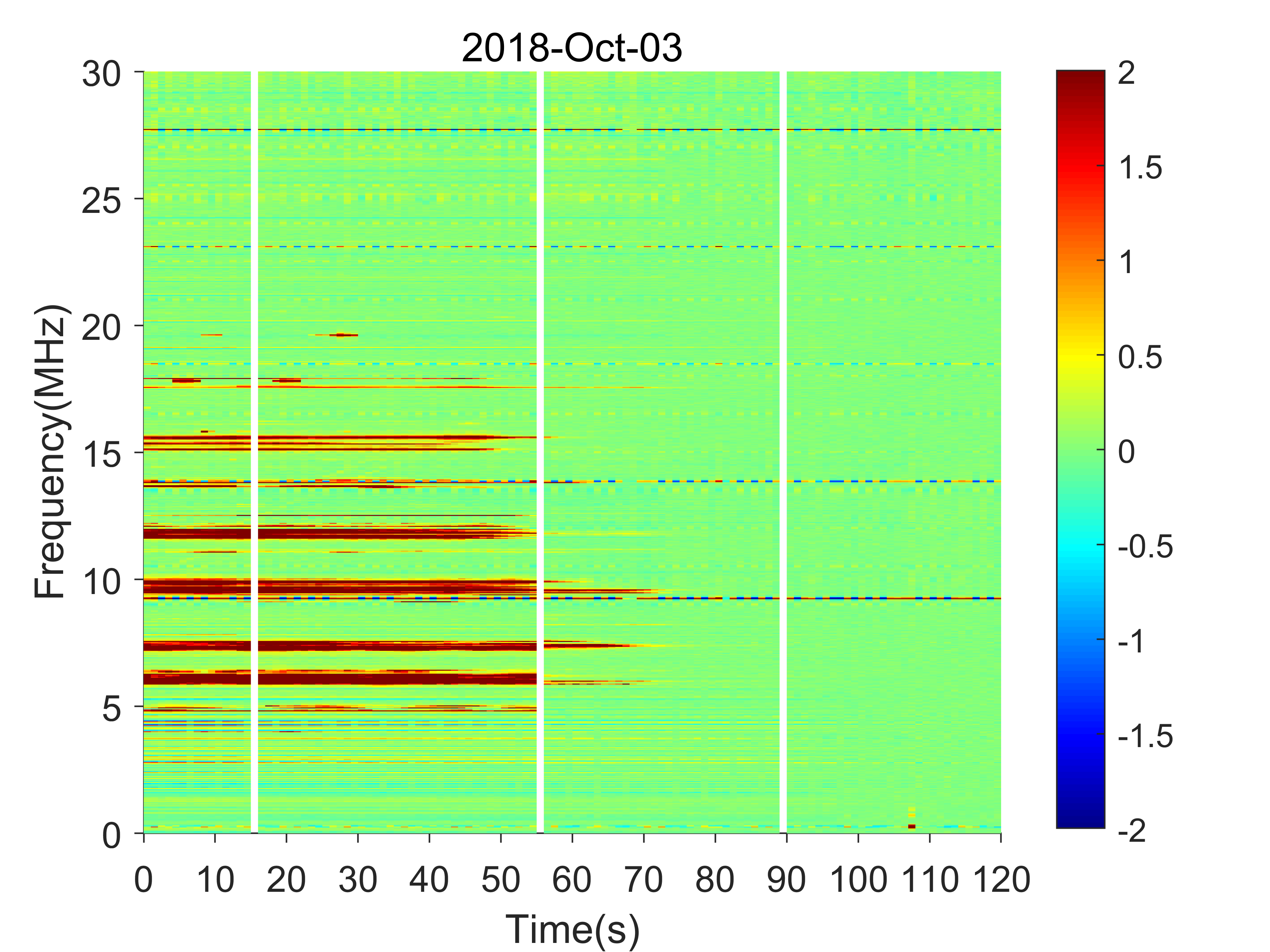}
    \end{minipage}
    \label{data-1003}
}

      \caption{The noise level detected by the LFIS receivers in the range of frequencies from 1 to 30 MHz (the vertical axis) on 2018 Oct 01 (panel (a)) and 2018 Oct 03 (panel (b)). Panel (a) shows egress and panel (b) ingress of the {\sl Longjiang-2} satellite with respect to the Lunar disc. The noise intensity is depicted in colour in the same scale as in the lower three panels of Fig.~\ref{spectrum}. The three white vertical lines in each plot show a start, medium phase (half a disc of Earth is visible from {\sl Longjiang-2)} and end of egress and ingress, respectively. The level of RFI noise increases visibly after egress, and decreases after the ingress. The asynchronous appearance of the noise at different spectral channels is due to the difference in the geographical location of the RFI sources on Earth. Narrow RFI signals at about 9.5, 14 and 28~MHz, which do not demonstarte dependence on the phases of Earth eclipse as see by {\sl Longjiang-2} are likely to have local origin in the satellite instrumentation. Qualitatively similar behaviour of the level of RFI has been detected in all egress and ingress LFIS observations.
      }
      \label{F:RFI-occ}
   \end{figure}

The advantage of Moon shielding of the Earth-originated RFI is further illustrated in Fig.~\ref{F:RFI-occ}. This illustration gives a clear qualitative evidence of the inhomogeneous distribution of the RFI over the spectrum. It is also reasonable to expect that the Earth-originated RFI depends on time. The effect of shielding of Earth-originated RFI should be exploited by future radio astronomy facilities on and near the Moon. Efforts for protecting the electromagnetic spectrum in the Moon's vicinity, and especially on its far side in the interests of radio astronomy should be considered as a high priority task. Certain activity in this topic has been recently re-activated by the International Academy of Astronautics\footnote{IAA Permanent Committee on Moon Farside Protection: https://iaaspace.org/about/permanent-committees/\#1658152007849-7e3d453e-d9ba, accessed on 2022.11.03.}.

%

\begin{figure}
   \centering
   \includegraphics[width=\hsize]{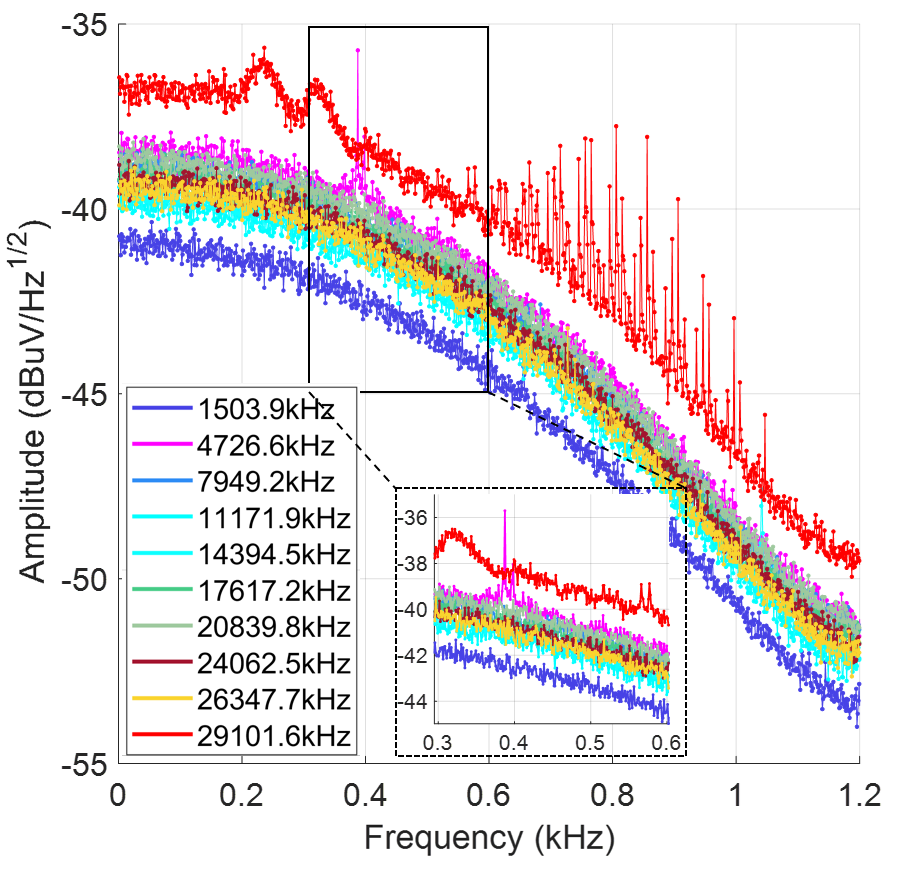}
      \caption{Spectra of 10 narrow-band channels of 1~kHz width. The spectra are created by Fourier transforms of the narrow band signals downloaded to the ground. The spectral resolution of the data is better than 1~Hz. A narrow-band RFI is present in the magenta curve (the 4.7~MHz channel) near 0.4~kHz, and a fine–structured RFI is clearly visible in the red curve (the 29~MHz channel) above 0.5~kHz.
              }
         \label{narrowband}
   \end{figure}

Additional complication of the Earth-originated RFI is its sporadic and time-variable character. Potential mitigation strategy of such the variable RFI might need to handle very narrow spectral channels, down to 1~Hz width as might be present in the spectrum shown in Fig.~\ref{narrowband}. Downlinking to the ground spectra of so high resolution will result in drastic escalation of requirements to the data rate of the communication systems. This challenge will have to be addressed in the future ULW missions with great attention.

As mentioned above, the receiver noise floor is rising visibly toward the low frequency end of the observed band (see Fig.~\ref{receiver-response}). After examining 10 narrow-band signals, provided by the filter bank designed originally for cross-correlation of signals from  {\sl Longjiang-1} and {\sl Longjiang-2} satellites, we found that the actual noise level across all narrow-band channels was almost flat. Fig.~\ref{narrowband} shows spectra of all narrow-band signals obtained with the antenna deployed in space. The frequency resolution of these spectra is better than 1~Hz. The spectrum of each 1~kHz channel presents a 3~dB slope which is consistent with the filter's response. The center frequencies of 10 signals cover a range from 1.5~MHz up to 29~MHz and are shown in the inset in Fig.~\ref{receiver-response}. The relative magnitudes of most channels are close to each other. It indicates that the intensity at different frequency channels are similar over the full band, thus the noise floor is approximately flat. Comparing the continuum spectrum provided by FFT(Fig.~\ref{receiver-response}) with the narrow band spectra in Fig.~\ref{narrowband}, we conclude that the frequency leakage from strong RFI raises the noise floor at lower frequency in Fig.~\ref{receiver-response}.  However, there are obvious RFI spikes in the 29.101~MHz channel (shown in red), and a very narrow spike in the 4.726~MHz channel (shown in magenta). We conclude that very high frequency resolution will be required in future ULW mission for mitigating narrow-band RFI spikes. 

%

\section{Conclusions}
\label{disc}

The LFIS experiment conducted the first ULW astronomical observation in the lunar orbit since the RAE-2 mission in the 1970s. It enabled in-flight demonstration of modern  technologies for prospective ULW missions. It also provided experimental verification of data calibration and processing. The LFIS observations conducted on the selenocentric orbit provided hitherto unavailable measurements of continuum emission covering the range from 1~MHz to 30~MHz. The LFIS instrumentation provided ultra-fine spectra with resolution $\sim$1~Hz within 10 narrow–band channels of the width of 1~KHz and of special value for RFI studies were observations conducted near the lunar limb transition. 

The LFIS observations allow us to formulate the following conclusions. We consider them as a set of inputs for designing future ULW missions, in particular, those which will address the original aim of LFIS as a demonstration of space-borne interferometry in the ULW range.

(1) The EMI control of a spacecraft is critical for ULW observations. Even though the noise floor determined onboard the spacecraft with conventional FFT of 39~kHz resolution was at the seemingly acceptable level, there were narrow-band RFI spikes  identified by the ultra-fine spectrum. Therefore, it is advisable to have ability to downlink digitized raw signal for in-depth high resolution spectrum analysis. This approach would help to apply most efficient RFI mitigation measures in future high sensitivity experiments. 

(2) We found that the terrestrial RFI is a few dB higher than the system temperature at the Earth-Moon distance. The required sensitivity for cosmological Dark Ages studies of about $10^{-2}$~K in the presence of RFI signal of the order of $10^{6}$~K necessitates an RFI suppression at least at the level of $-$80~dB. The LFIS observations demonstrated sharp drop in the RFI level after {\sl Longjiang-2} transition through the lunar limb from the visible segment of the orbit to the occulted one. This effect is slightly asynchronized at different spectral channels due to different locations of the RFI sources over the Earth surface. 

(3) The LFIS experiment demonstrated the value of RFI monitoring over the near and far sides of the Moon (i.e., at the visible and occulted arcs of the selenocentric orbit). Such the monitoring enabled us to disentangle terrestrial and local instrumental RFI. A simple subtraction algorithm allowed us to mitigate the satellite-originated RFI by about 30~dB. Such the approach would be beneficial for future ULW missions, but will require further refinements for deeper RFI mitigation, including narrow-band analysis of the signal and its polarization properties.

LFIS measurements are also potentially applicable to derive ULW global spectrum with 10 narrow band signals, but a sophisticated system model needs to be created for quantitative analysis of similar data in the future missions.

\begin{acknowledgements}
We are grateful to the Lunar Exploration and Space Program Center, China National Space Administration for the piggybacking opportunity for the {\sl Longjiang} flight on the Chang'E-4 Lunar mission. The China-Europe joint team for the DSL concept helped to refine many details of the experiment presented in this paper. We acknowledge with gratitude very useful comments provided by the anonymous reviewer of the manuscript of this paper.
\end{acknowledgements}

\paragraph{Data availability}

The datasets analyzed during the current study are available from the National Space Science Data Center, owned by the Lunar Exploration and Space Program Center, China National Space Administration but restrictions apply to the availability of these data, which were used under license from the Lunar Exploration and Space Program Center, and so are not publicly available. Data are however available from the authors upon reasonable request and permission of the Lunar Exploration and Space Program Center.

\paragraph{Conflict of interest}
The authors have no competing interests to declare that are relevant to the content of this paper.

%
%

\bibliography{ULFISbibfile}   

\end{document}